\def\papertitle{A Dual-Mode Faust-to-CLAP Compilation System}
\def\paperauthorA{Facundo Franchino}
\def\paperauthorB{St\'{e}phane Letz}
\def\paperauthorC{Jatin Chowdhury}

\documentclass[twoside,a4paper]{article}
\usepackage{IFC-26}
\usepackage{amsmath,amssymb,amsfonts,amsthm}
\usepackage{euscript}
\usepackage[utf8]{inputenc}
\usepackage[T1]{fontenc}
\usepackage{ifpdf}
\usepackage{algorithm}
\usepackage{algpseudocode}
\usepackage{tikz}
\usetikzlibrary{shapes.geometric, arrows.meta, positioning}

\usepackage{color}
\usepackage{listings}
\definecolor{mygrey}{rgb}{0.96,0.96,0.96}
\definecolor{mygreen}{rgb}{0,0.5,0}
\definecolor{myblue}{rgb}{0,0,0.7}
\usepackage[english]{babel}
\usepackage{caption}
\usepackage{subfig, color}

\ninept

\usepackage{times}

\usepackage{graphicx}
\usepackage{float}
\usepackage[
  pdftitle={\papertitle},
  pdfauthor={\paperauthorA, \paperauthorB, \paperauthorC},
  colorlinks=false,
  bookmarksnumbered,
  pdfstartview=XYZ
]{hyperref}
\usepackage[figure,table]{hypcap}

\newcommand{\F}{\textsc{Faust}}
\newcommand{\CLAP}{\textsc{Clap}}

\title{\papertitle}

\threeaffiliations{
\paperauthorA \,\sthanks{This work was supported by Google Summer of Code 2025.}}
{School of Physics, Engineering and Technology \\ University of York, United Kingdom \\ \href{mailto:facundof@mit.edu}{\textbf{facundof@mit.edu}}}
{\paperauthorB}
{\href{http://grame.fr}{GRAME-CNCM} \\ Lyon, France \\ \href{mailto:letz@grame.fr}{\textbf{letz@grame.fr}}}
{\paperauthorC}
{Department of Electrical Engineering and Computer Science \\ Massachusetts Institute of Technology, USA \\ \href{mailto:chowdsp@mit.edu}{\textbf{chowdsp@mit.edu}}}

\begin{document}
\DeclareGraphicsExtensions{.png,.jpg,.pdf}
\pagestyle{fancy}
\cfoot{\thepage}

\maketitle

\begin{abstract}
We describe \texttt{faust2clap}, a framework establishing the first officially maintained compilation pathway from \F{} DSP specifications to the \CLAP{} format. The system operates in two different modes. A static mode employs ahead-of-time compilation to yield native binaries of optimal efficiency, while a dynamic mode uses runtime interpretation to permit DSP code modification without interrupting the host application. This latter capability addresses a persistent friction in audio software development, namely the cumulative overhead of the edit, compile, and reload cycle.

We detail the algorithmic machinery underlying both modes, focusing specifically on the problem of parameter identity. To preserve both parameter values and their bindings to host automation across structural DSP mutations, we introduce an address-based identity matching algorithm and a stable slot allocation scheme. The implementation, comprising approximately 2,400 lines of C++ architecture and Python tooling code, has been integrated into the main \F{} distribution.
\end{abstract}

\section{Introduction}
\label{sec:intro}

The practice of audio plugin development presents a characteristic tension between efficiency and iteration. Real-time audio processing demands computational efficiency, which suggests native compilation and careful optimisation. Conversely, the creative refinement of signal processing algorithms calls for immediate feedback and rapid experimentation. 

Traditional workflows resolve this tension decisively in favour of efficiency. A developer writes code, compiles it, loads the resulting binary into a host application, evaluates the behaviour, and returns to the editor. Each cycle consumes time, and this overhead may accumulate significantly. 

The \F{} programming language \cite{Orlarey2009} mitigates this difficulty through its functional approach to DSP specification. Rather than managing sample buffers and loop indices, the programmer describes signal transformations algebraically. From this representation, the compiler then generates the efficient imperative code. Yet, the fundamental workflow remains bound to the compilation and reload cycle.

The \CLAP{} standard \cite{CLAP2022}, introduced in 2022, represents the current state of the art in plugin interface design. Its pure C specification makes it binary-compatible across compilers, its threading model is explicit, and its licence carries no patent encumbrances.

This paper describes \texttt{faust2clap}, a framework bridging these technologies. Our principal contributions are:

\begin{enumerate}
\item A direct compilation pathway from \F{} to \CLAP{}.
\item A dynamic mode permitting runtime DSP modification without host restart.
\item An algorithm for stable parameter mapping that preserves host automation across reloads.
\item Automatic heuristics for classifying DSPs as effects or instruments.
\end{enumerate}

\section{Background and Related Work}
\label{sec:background}

The \F{} ecosystem includes architecture files for VST, Audio Unit, and LV2 targets, \cite{FaustArch}. These produce statically compiled plugins through a standard pipeline. Our work extends this ecosystem with \CLAP{} support and introduces a dynamic runtime mode.

The Heavy Compiler Collection \cite{HVCC} provides analogous functionality for Pure Data patches, producing ahead-of-time compiled binaries similar to our static mode. Commercial tools such as RNBO \cite{RNBO} enable Max/MSP patch export to various formats, including \CLAP{}, but operate within a proprietary ecosystem and require full recompilation for deployment.

Camomile \cite{Guillot2018,GuillotJIM2018} embeds a Pure Data interpreter within a plugin shell, restoring state by parameter index rather than semantic identity; structural patch modifications therefore invalidate host automation mappings. Amati \cite{Amati2022} provides runtime \F{} compilation for VST3 and Audio Unit targets with a similar index-based approach. The \texttt{faust2clap} framework introduces address-based identity matching that preserves automation across structural changes, provides MIDI support for instrument synthesis, and offers both static compilation for production deployment and dynamic interpretation for rapid iteration, all integrated into the mainline \F{} distribution.

\section{Architecture}
\label{sec:architecture}

The framework implements two distinct compilation pathways unified by a common parameter management infrastructure.

\subsection{The Static Pathway}

The static compilation pathway transforms a functional specification directly into a native machine code binary. The Python orchestrator invokes the \F{} compiler alongside the \texttt{clap-arch.cpp} architecture file, which serves as an adapter between the generated \texttt{dsp} interface and the \CLAP{} API. The concatenated C++ code is subsequently compiled and linked via CMake. Because this mode produces native machine code, audio processing executes at full theoretical speed with zero runtime interpretation overhead.

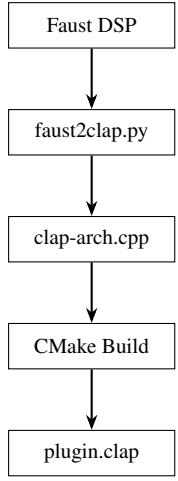
\begin{figure}[ht]
\centering
\begin{tikzpicture}[
    node distance=0.8cm,
    box/.style={rectangle, draw, minimum width=2.2cm, minimum height=0.6cm, align=center, font=\footnotesize},
    arrow/.style={-{Stealth[length=2mm]}, thick}
]
\node[box] (dsp) {Faust DSP};
\node[box, below=of dsp] (faust2clap) {faust2clap.py};
\node[box, below=of faust2clap] (arch) {clap-arch.cpp};
\node[box, below=of arch] (cmake) {CMake Build};
\node[box, below=of cmake] (plugin) {plugin.clap};

\draw[arrow] (dsp) -- (faust2clap);
\draw[arrow] (faust2clap) -- (arch);
\draw[arrow] (arch) -- (cmake);
\draw[arrow] (cmake) -- (plugin);
\end{tikzpicture}
\caption{\label{fig:static_workflow}{\it Static compilation workflow. The Python orchestrator invokes the Faust compiler with the CLAP architecture file, generating C++ source that CMake compiles into a native plugin binary.}}
\end{figure}

\vspace{-2mm}
\subsection{The Dynamic Pathway}

\noindent\begin{minipage}{\columnwidth}
\centering
\begin{tikzpicture}[
    node distance=0.8cm,
    box/.style={rectangle, draw, minimum width=2.2cm, minimum height=0.6cm, align=center, font=\footnotesize},
    arrow/.style={-{Stealth[length=2mm]}, thick}
]
\node[box] (dsp) {Faust DSP};
\node[box, below=of dsp] (efsw) {efsw watch};
\node[box, below=of efsw] (libfaust) {libfaust};
\node[box, below=of libfaust] (bytecode) {VM bytecode};
\node[box, below=of bytecode] (audio) {Live audio};

\draw[arrow] (dsp) -- (efsw);
\draw[arrow] (efsw) -- (libfaust);
\draw[arrow] (libfaust) -- (bytecode);
\draw[arrow] (bytecode) -- (audio);
\end{tikzpicture}

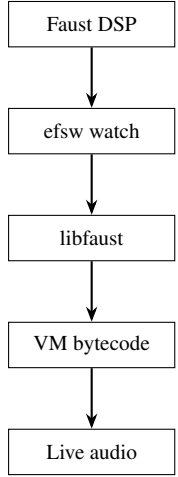
\captionof{figure}{\label{fig:dynamic_workflow}{\it Dynamic hot-reload workflow. The efsw library \cite{efsw} monitors the DSP source file; upon modification, libfaust recompiles to VM bytecode without interrupting the host.}}
\end{minipage}

\vspace{2mm}
The dynamic pathway employs the \texttt{libfaust} interpreter backend for runtime compilation. A dedicated thread monitors the file system for modifications to the target DSP file. Upon detecting a change, the framework invokes the interpreter factory to generate updated bytecode. The resulting bytecode executes upon a stack-based virtual machine, which maintains strictly separate integer and floating-point memory heaps for deterministic state management. This process permits the modification of DSP code without triggering a host restart.

\section{Algorithms and Parameter Identity}
\label{sec:algorithms}

When a user modifies and reloads a DSP specification, the internal parameter ordering shifts. A naive reassignment of parameter indices would instantly invalidate the automation data established by the host workstation. We decompose this problem into two tasks: preserving values via address-based identity matching, and assigning stable automation identifiers.

\subsection{Address-Based Identity Matching}

Each \F{} parameter possesses a hierarchical address string, such as \texttt{/Reverb1/Damp}, reflecting its position in the user interface hierarchy. Upon a hot-reload, Algorithm \ref{alg:preservation} builds a hash map from old addresses to current values, then iterates through the new DSP. If a parameter's address exists in the map, its value is clamped to the new bounds and restored.

\begin{algorithm}[ht]
\caption{Parameter Value Preservation}
\label{alg:preservation}
\begin{algorithmic}[1]
\Require Old DSP instance $D$, new DSP instance $D'$
\Ensure Values preserved where addresses match
\State $\mathcal{V} \gets \emptyset$ \Comment{Address $\to$ value map}
\For{$i \gets 0$ \textbf{to} $D.\texttt{paramCount}() - 1$}
    \State $\mathcal{V}[D.\texttt{getAddress}(i)] \gets D.\texttt{getValue}(i)$
\EndFor
\For{$j \gets 0$ \textbf{to} $D'.\texttt{paramCount}() - 1$}
    \State $a \gets D'.\texttt{getAddress}(j)$
    \If{$a \in \mathcal{V}$}
        \State $v \gets \texttt{clamp}(\mathcal{V}[a], D'.\texttt{getMin}(j), D'.\texttt{getMax}(j))$
        \State $D'.\texttt{setValue}(j, v)$
    \EndIf
\EndFor
\end{algorithmic}
\end{algorithm}

This algorithm guarantees that if a parameter's address string remains identical across reloads, its semantic value is preserved. If the developer explicitly renames the path in the source code, the link is severed and the parameter reverts to its initialisation default. 

\subsection{Stable Slot Mapping}

To preserve host automation bindings, we allocate a fixed array of $k$ constant slots (currently $k = 12$). This design follows the precedent established by Surge XT Effects \cite{SurgeXT}, which employs a similar fixed-slot architecture to expose multiple effect types through a single plugin instance. The value was chosen as a practical upper bound for effects, which typically expose fewer than a dozen controls; the \CLAP{} API requires stable parameter identifiers, and dynamic slot allocation would necessitate parameter rescanning that breaks automation bindings. While the mapping from these slots to \F{} parameter indices may vary across reloads, the slot identifiers exposed to the host remain immutable.

The reconstruction proceeds in two passes. First, we traverse existing slots. If the address previously mapped to a slot survives in the new DSP, that binding is preserved and the slot points to the parameter's new index. Second, we identify novel parameters (those not yet bound to any slot) and assign them to free slots in order.

This logic guarantees that host automation curves remain bound to the correct variable, provided the address string persists. If the compiled DSP unit declares more than 12 parameters, the mapping truncates deterministically; excess parameters process correctly internally but are not exposed to host automation. 

\subsection{DSP Classification Heuristics}

Automatic classification enables the framework to define proper polyphonic allocations without user intervention. The orchestrator utilises a three-tier heuristic. First, it parses the source code for explicit \texttt{declare} metadata. Because developers frequently omit metadata, the second tier performs lexical analysis on the filename for common keywords: effect indicators include \texttt{reverb}, \texttt{delay}, \texttt{filter}, \texttt{compress}, \texttt{phaser}, and \texttt{chorus}; instrument indicators include \texttt{synth}, \texttt{organ}, \texttt{piano}, and \texttt{osc}. The third tier performs structural analysis by compiling the C++ geometry and querying \texttt{getNumInputs()}. A strict zero dictates a pure generator (instrument), while any value greater than zero dictates a processor (effect). In edge cases where an instrument requires audio input, such as a vocoder, the user must explicitly define the behaviour via tier-one metadata to override the structural heuristic.

\section{Implementation Details}
\label{sec:implementation}

The hot-reload mechanism coordinates two threads, a file watcher and the audio processing thread. The file watcher observes the disk and sets an atomic flag upon detecting a file modification. The audio thread checks this flag and, when asserted, performs inline recompilation via the interpreter. This design prioritises architectural simplicity and predictable reload semantics over uninterrupted playback; 
a brief discontinuity occurs during recompilation, typically lasting 5--60ms depending on DSP complexity. While this violates conventional real-time audio constraints \cite{Bencina2011}, the trade-off is acceptable for a development-oriented workflow tool where rapid iteration takes precedence over seamless audio during code changes.

Host audio pipelines may supply 32-bit or 64-bit floating-point buffers. The framework detects these at runtime and performs format conversion as needed.

The static pathway supports optional polyphonic synthesis via the \texttt{mydsp\_poly} wrapper, allocating 16 voices by default for DSPs classified as instruments. The dynamic pathway currently operates in monophonic mode only; polyphonic hot-reload would require preserving per-voice state across recompilations, which remains future work.

\section{Evaluation}
\label{sec:evaluation}

\begin{figure}[ht]
\centering
\includegraphics[width=1.12\columnwidth]{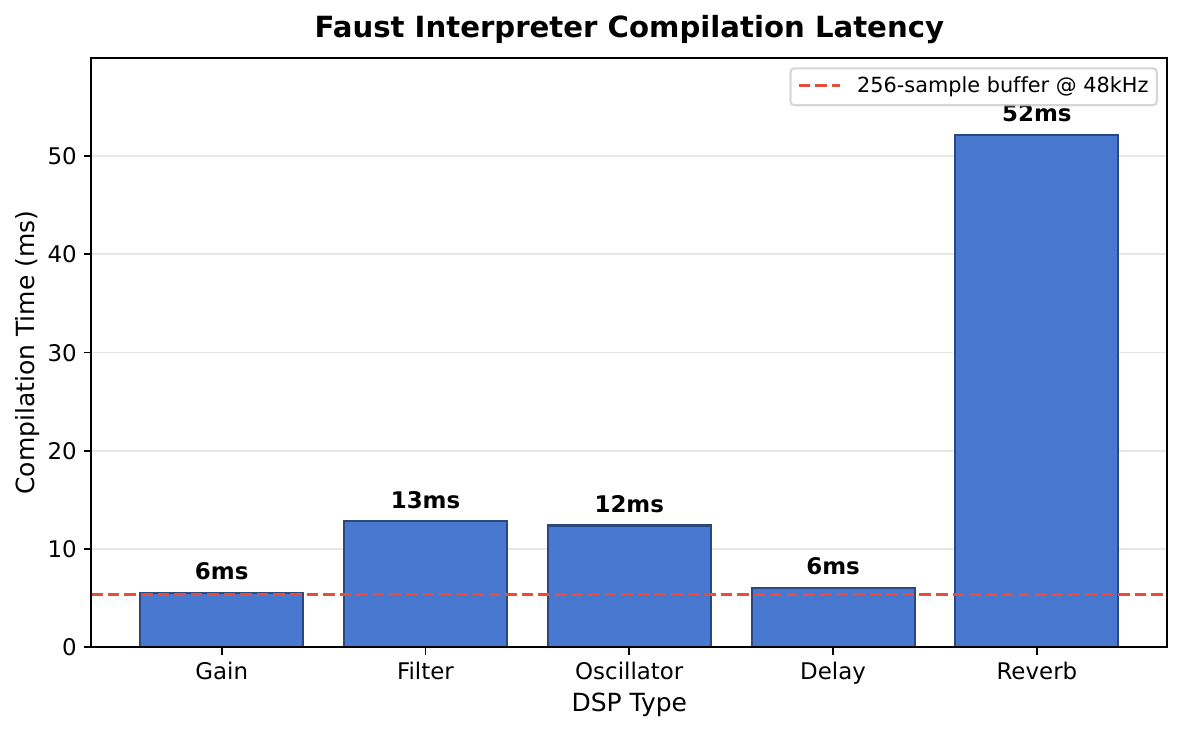}
\caption{\label{fig:latency}{\it Interpreter compilation latency for representative DSP topologies (mean of 10 iterations, variance below 5\%). The dashed line shows a single 256-sample buffer duration (5.3ms) for scale; compilation is a one-time event per reload, not a recurring deadline. All tested DSPs compile within 60ms, acceptable for interactive development.}}
\end{figure}

To validate the parameter identity machinery, we measured reload latency across DSP files of varying complexity and verified the preservation guarantees. All measurements were conducted on an Apple M2 processor running macOS 26.3, using REAPER 7.0 as the host application. Figure \ref{fig:latency} illustrates compilation time as a function of DSP complexity, averaged over 10 iterations per DSP topology.

Table \ref{tab:benchmarks} summarises the preservation guarantees.

\begin{table}[ht]
  \caption{\itshape Parameter identity validation. Value preservation indicates whether parameter values survived reload when addresses were unchanged. Slot persistence indicates whether host automation bindings remained valid.}
    \centering
    \begin{tabular}{|l|c|c|c|}
        \hline
        \textbf{DSP Type} & \textbf{Latency (ms)} & \textbf{Value} & \textbf{Slot} \\\hline
        Gain & 6 & \checkmark & \checkmark \\
        Filter & 13 & \checkmark & \checkmark \\
        Oscillator & 12 & \checkmark & \checkmark \\
        Delay & 6 & \checkmark & \checkmark \\
        Reverb & 52 & \checkmark & \checkmark \\\hline
    \end{tabular}
    \label{tab:benchmarks}
\end{table}

As expected, renaming a parameter address in the source code severs continuity. The old value is discarded and the parameter initialises to its default. This behaviour is intentional and documented in Section \ref{sec:algorithms}.

\begin{figure}[ht]
\centering
\includegraphics[width=1.05\columnwidth, height=6cm]{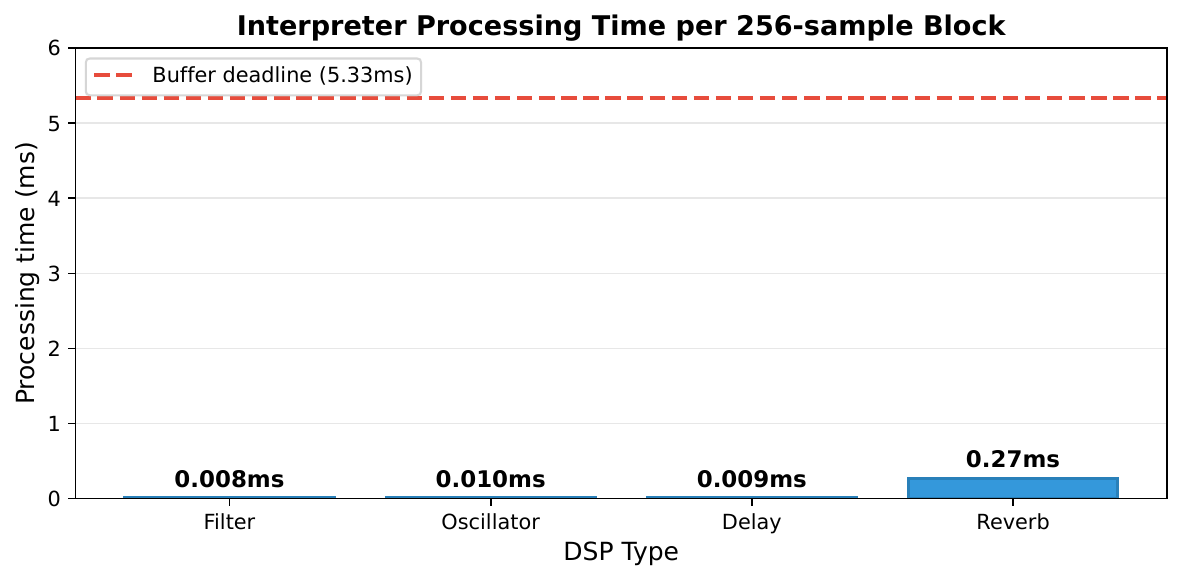}
\caption{\label{fig:interp_perf}{\it Interpreter processing time per 256-sample block at 48kHz. The dashed line indicates the real-time deadline; all DSPs complete well below this threshold.}}
\end{figure}

Figure \ref{fig:interp_perf} and Table \ref{tab:cpu} present interpreter performance for representative DSP topologies. The critical metric is per-block processing time relative to the 5.33ms deadline imposed by a 256-sample buffer at 48kHz. Even the most demanding topology, a stereo reverb comprising multiple allpass chains, completes in 0.27ms, providing 20$\times$ headroom. Simpler topologies complete in under 0.02ms.

\begin{table}[ht]
  \caption{\itshape Interpreter processing time per block. Headroom indicates how many identical instances could run simultaneously before exceeding real-time.}
    \centering
    \begin{tabular}{|l|c|c|}
        \hline
        \textbf{DSP Type} & \textbf{Time/block} & \textbf{Headroom} \\\hline
        Filter (resonant LP) & 0.008ms & 666$\times$ \\
        Oscillator (sine) & 0.010ms & 533$\times$ \\
        Delay (feedback) & 0.009ms & 592$\times$ \\
        Reverb (\texttt{dm.zita\_light}) & 0.27ms & 20$\times$ \\\hline
    \end{tabular}
    \label{tab:cpu}
\end{table} 

\section{Conclusions}
\label{sec:conclusions}

We have presented \texttt{faust2clap}, establishing the primary integration of the \CLAP{} standard into the \F{} ecosystem. The framework satisfies both the deployment requirement for high-efficiency static binaries and the developer requirement for frictionless, interactive iteration. 

Our core contribution is a CLAP backend for Faust and the stable slot mapping logic. The latter which, unlike index-based approaches used by comparable tools, preserves host automation data across structural DSP mutations provided parameter addresses remain unchanged. Without this mechanism, dynamic reloading would be actively hostile to modern production workflows.

The static compilation pathway has been validated on macOS, with Linux support requiring minor build system adjustments for local library paths. The dynamic interpreter architecture is currently restricted to macOS due to platform-specific file-watching and bundle-loading dependencies; Linux support is under active development.

Future work includes two directions. First, investigating LLVM backend integration, which is expected to reduce the interpreter overhead penalty significantly. Second, refactoring the hot-reload mechanism to perform compilation on a background thread with double-buffered DSP instances, eliminating the brief discontinuity during reloads and adhering to established real-time audio programming practices \cite{Bencina2011}.

\noindent Source code: \url{https://github.com/cucuwritescode/faust2clap}

\section{Acknowledgements}

This work was supported by Google Summer of Code 2025. We thank the \CLAP{} community for standardisation efforts, and the \F{} team at GRAME for maintaining the ecosystem upon which this work depends.

\bibliographystyle{IEEEbib}
\bibliography{IFC-26}

\begin{thebibliography}{10}

\bibitem{Orlarey2009}
Yann Orlarey, St{\'e}phane Letz, and Dominique Fober,
\newblock ``{FAUST}: an efficient functional approach to {DSP} programming,''
\newblock in {\em New Computational Paradigms for Computer Music}. Delatour, Paris, France, 2009.

\bibitem{CLAP2022}
{Free Audio Community},
\newblock ``{CLAP}: {CL}ever {A}udio {P}lug-in {API},'' \url{https://github.com/free-audio/clap}, 2022,
\newblock Version 1.0.0 released June 2022.

\bibitem{FaustArch}
St{\'e}phane Letz, Yann Orlarey, and Dominique Fober,
\newblock ``Faust architecture files,'' \url{https://faustdoc.grame.fr/manual/architectures/}, 2024.

\bibitem{HVCC}
{Wasted Audio},
\newblock ``Heavy compiler collection (hvcc),'' \url{https://wasted-audio.github.io/hvcc/}, 2021.

\bibitem{RNBO}
{Cycling '74},
\newblock ``{RNBO} audio plugin export,'' \url{https://rnbo.cycling74.com/}, 2023.

\bibitem{Guillot2018}
Pierre Guillot,
\newblock ``Camomile: Creating audio plugins with {Pure Data},''
\newblock in {\em Proceedings of the Linux Audio Conference (LAC-18)}, Berlin, Germany, 2018.

\bibitem{GuillotJIM2018}
Pierre Guillot,
\newblock ``Camomile, enjeux et d{\'e}veloppements d'un plugiciel audio embarquant {Pure Data},''
\newblock in {\em Journ{\'e}es d'Informatique Musicale (JIM 2018)}, Amiens, France, May 2018.

\bibitem{Amati2022}
Gr{\'e}goire Locqueville,
\newblock ``{Amati}: A {VST} plugin for live-coding in {Faust},'' \url{https://github.com/glocq/Amati}, 2022.

\bibitem{efsw}
Mart{\'i}n~Lucas Golini,
\newblock ``{efsw}: Entropia file system watcher,'' \url{https://github.com/SpartanJ/efsw}, 2024.

\bibitem{SurgeXT}
{Surge Synth Team},
\newblock ``{Surge XT Effects},'' \url{https://surge-synth-team.org/surge-effects/}, 2024.

\bibitem{Bencina2011}
Ross Bencina,
\newblock ``Real-time audio programming 101: Time waits for nothing,'' 2011.

\end{thebibliography}

\end{document}